\documentclass[twocolumn,showpacs,preprintnumbers,amsmath,amssymb]{revtex4}
\usepackage{graphicx}% Include figure files
\usepackage{dcolumn}% Align table columns on decimal point
\usepackage{bm}% bold math

\begin{document}
%************************************************************************
\title{Searching for the statistically equilibrated systems formed in
heavy ion collisions}
\author{Al. H. Raduta$^{1,2}$, Ad. R. Raduta$^{1,2}$}
\affiliation{
        $^1$GSI, D-64220 Darmstadt, Germany\\
        $^2$NIPNE, RO-76900 Bucharest, Romania}

\begin{abstract}
  Further improvements and refinements are brought to the
  microcanonical multifragmentation model [Al. H. Raduta and Ad. R.
  Raduta, Phys. Rev. C {\bf 55}, 1344 (1997); {\it ibid.} {\bf 61},
  034611 (2000)].  The new version of the model is tested on the
  recently published experimental data concerning the Xe+Sn at 32
  MeV/u and Gd+U at 36 MeV/u reactions. A remarkable good simultaneous
  reproduction of fragment size observables and kinematic
  observables is to be noticed. It is shown that the equilibrated
  source can be unambiguously identified.
\end{abstract}
\maketitle

\section{Introduction}

It is known from more than 15 years that violent heavy ion collisions
lead to an advanced disassembly of the compound system known under the
name of nuclear fragmentation.  The asymptotically resulted fragments
are the only experimental link with the primordial process. Good
agreements between various observables related to these fragments and
results of various models assuming uniform population of the system
phase space \cite{Gross1,Bondorf,Randrup} led to the conclusion that
a statistically equilibrated nuclear source is at the origin of the
fragmentation process. The source size, its excitation energy and its
volume are thus quantities which can only be indirectly evaluated by
comparisons between experimental data and statistical
multifragmentation models predictions. The comparison process is
complicated by various quantities polluting the pure statistical
signals such as preequilibrium particle emission, collective radial
expansion, Coulomb propagation of the break-up fragments, secondary
particle emissions.  A complete theoretical reproduction of
experimental fragmentation event topology (never achieved so far)
would help to answer questions fundamental for determining the
thermodynamics taking place in such systems. For example, the
identification of the experimental freeze-out volume would allow to
locate the process in the phase diagram and, therefore, would help
solving a problem which made the subject of a strong
debate over the last decade: is there any phase transition taking
place in nuclear matter?

The aim of the present paper is to investigate whether an unambiguous
determination of the experimental statistically equilibrated source
can be achieved by means of the microcanonical model from Ref.
\cite{noi1,temp3}. To this aim, the model is refined by including the
conservation of the angular momentum and completed with a Coulomb
propagation stage of the primary fragments with the possibility of
superimposing radial flow.  A complete set of experimental
fragmentation data concerning the reactions Xe+Sn at 32 MeV/u and Gd+U
at 36 MeV/u recently published in Ref. \cite{indra1,indra2} is used
for comparison with the model results. The paper is structured as
follows: Section II gives a brief presentation of the employed
microcanonical model and describes its new refinements and
improvements. In Section III a comparison between the model prediction
and the above mentioned experimental data is presented. The influence
of various model parameters on different observables is discussed in
Section IV.  Conclusions are drawn in Section V.

\section{The model: brief review and improvements}

The microcanonical multifragmentation model used in the present study
was proposed in Ref. \cite{noi1,temp3}. The modifications from the
previous version of the model concern the primary break-up stage,
momentum generation, inclusion of collective radial expansion, Coulomb
propagation, and the secondary emission stage.\\
{\em i) Break-up stage}\\
This stage concerns the disassembly of a statistically equilibrated
nuclear source $(A,~Z,~E,~V)$ (the mass number, the atomic number, the
excitation energy and the freeze-out volume respectively). The basic
assumption of the model is equal probability between all
configurations $C:\{A_i,~Z_i,~\epsilon_i,~{\bf r}_i,~{\bf
  p}_i,~~i=1,\dots,N\}$ (the mass number, the atomic number, the
excitation energy, the position and the momentum of each fragment $i$
of the configuration $C$, composed of $N$ fragments).  Fragments are
assumed to be spherical, are not allowed to overlap each other and are
placed into a spherical recipient of volume $V$. In the previous
version of the model \cite{noi1} the system was considered subject to
the standard microcanonical constraints: $\sum_i A_i=A$, $\sum_i
Z_i=Z$, $\sum_i {\bf p}_i={\bf P}$ (=0 in the c.m. frame), $E$ -
constant. An extra constraint will be considered herein: the
conservation of the angular momentum, $\sum_i {\bf r}_i \times {\bf
  p}_i={\bf L}$.  The integration over fragment momenta can be
analytically performed subject to the above mentioned constraints:
\begin{widetext}
\begin{eqnarray}
&&\int\prod_{i=1}^N {\rm d} {\bf p}_i~\delta\left(H-E\right)~ 
        \delta\left(\sum_i {\bf p}_i -{\bf P}\right)~
        \delta\left(\sum_i {\bf r}_i \times {\bf p}_i-{\bf L}\right)\nonumber\\
        &=&\frac{2\pi}{\Gamma\left(\frac32(N-2)\right)}
        \left(\frac{\prod_i m_i}{\sum_i m_i}\right)^{3/2}
        \frac1{\sqrt{\det {\bf I}}}
        \left[2\pi 
        \left(K-\frac{P^2}{2M}-\frac12{\bf L}^T {\bf I}^{-1} {\bf L}\right)
        \right]^{\frac32(N-2)-1}
        \label{eq:int}
\end{eqnarray}
\end{widetext}
Here $M=\sum_i m_i$, ${\bf I}$ is the inertial tensor of the system:
$I_{\alpha\beta}=\sum_i m_i \left(r_i^2
  \delta_{\alpha\beta}-r_i^{\alpha}r_i^{\beta}\right)$ with
$\alpha,\beta=1,2,3$, ${\bf L}^T {\bf I}^{-1} {\bf L}=
\sum_{\alpha\beta=1}^3 L_{\alpha} I_{\alpha\beta}^{-1} L_{\beta}$,
$H=\sum_i p_i^2/(2 m_i)+\sum_{i<j}V_{ij}+\sum_i\epsilon_i-\sum_i B_i$
and $K=E-\sum_{i<j}V_{ij}-\sum_i \epsilon_i+\sum_i B_i$ ($V_{ij}$
stands for the Coulomb interaction between fragments $i$ and $j$). Of
course, the model corresponds to the c.m. frame (i.e. ${\bf r}_i$ with
$i=1,\dots,N$ are positions in the c.m. frame) where ${\bf P}=0$
\footnote{For each particular fragmentation event the origin of the
  system is considered in its center of mass and not in the
  center of the spherical ``recipient'' frame.  Therefore, at each
  step of the Metropolis type simulation one has to consider the
  transformation ${\bf r}_i={\bf r}_0^i-{\bf r}_{\rm c.m.}$,
  $i=1,\dots,N$. Here ${\bf r}_0^i$, $i=1,\dots,N$ represent the
  positions of the fragments subject to the origin of the spherical
  recipient, generated as in \cite{noi1}.}. If one also imposes
${\bf L}=0$ (which is the hypothesis employed in the present calculations) 
then the last factor in eq. (\ref{eq:int}) becomes $(2\pi
K)^{3/2N-2}$.  The only modifications brought by the inclusion of the
angular momentum conservation constraint to the statistical weight of
a configuration $C':\{A_i,~Z_i,~\epsilon_i,~{\bf r}_i,~~i=1,\dots,N\}$
($W_{C'}$) from Ref. \cite{noi1} are the inclusion of the factor
$1/\sqrt{\det {\bf I}}$ and the replacement of $N$ with $N-1$.  These
new weights can be employed in a Metropolis-type simulation which
allows the determination of the average value of any system observable
in the very same manner as in Ref. \cite{noi1}.

Fragments with $A \le 4$ are considered without excitation degrees of
freedom except for the $\alpha$ particle for which the few levels
larger than 20 MeV with $\Gamma \le 2.01$ MeV have been considered. 
These fragments are weighted in $W_{C'}$ with their energy levels
degeneracies. Larger fragments carry internal excitation. For them, the
following level density formula is included in the statistical weight 
of a configuration $C'$ (see Ref. \cite{temp3}):
\begin{equation}
\rho(\epsilon)=\frac{\sqrt{\pi}}{12~a^{1/4}\epsilon^{5/4}}
\exp(2 \sqrt{a\epsilon})\exp(-\epsilon/\tau)
\end{equation}
with $a=0.114 A+0.098 A^{2/3}$ MeV$^{-1}$ \cite{iljinov} 
and $\tau=9$ MeV. The factor
$\exp(-\epsilon/\tau)$ is introduced to account for the dramatic
decrease of the excited levels lifetime at high excitation 
energies \cite{Randrup}.\\
{\em ii) Generation of the primary decay fragment momenta}\\ 
While
integration over the fragments' momenta has been carried out in order
to simplify the Metropolis simulation, to produce events similar to
the experimental ones one has to generate momenta for each given
configuration $C'$. Each of these events is characterized by a 
kinetic energy $K$. Therefore, we have to deal with the following
computational task: generate uniformly momenta for a system composed
of $N$ particles such that $\sum_i p_i^2/2m_i=K$, $\sum_i {\bf
p}_i=0$ and $\sum_i {\bf r}_i\times{\bf p}_i={\bf L}$. This problem
was solved in an elegant way in Ref. \cite{randrup}. There it is
proved that the following generation gives the right sampling of the
system momenta, in agreement with the above mentioned constraints:
Pick a preliminary set of $N$ particle momenta from a arbitrary
canonical distribution, then eliminate the overall translational and
rotational motion by making the transformation ${\bf p}_i \rightarrow
{\bf p}_i-m_i ({\bf P'}/M+\mbox{\boldmath $\omega'$}\times {\bf r}_i)$ (where
${\bf P'}=\sum_i {\bf p}_i$ and $\mbox{\boldmath $\omega'$}=\sum_i {\bf r}_i \times {\bf p}_i
\cdot {\bf I}^{-1}$), then spin the system such as to match the
desired angular momentum ${\bf L}$: ${\bf p}_i \rightarrow {\bf
p}_i+m_i~\mbox{\boldmath $\omega$}\times {\bf r}_i)$ with 
$\mbox{\boldmath $\omega$}={\bf L}^T
\cdot {\bf I}^{-1}$ and finally, renormalize the momenta such as to
match the available energy $K-\frac12~\mbox{\boldmath $\omega$} 
\cdot {\bf L}$.\\
{\em iii) Radial flow}\\ 
After generating the fragment momenta
corresponding to the primary decay, inclusion of nonequilibrium
effects such as collective radial expansion (flow) can be easily
superimposed.  Here we use the following parameterization for the flow velocity
of fragment $i$: ${\bf v}_{f}^i={\bf v}_0~(r_i/R)^{\alpha}$, with
${\bf v}_0=v_0({\bf r}_i/r_i)$ and $\alpha$ a real number defining the
flow profile. After including the radial flow the momentum of the
$i$th fragment reads: ${\bf p}_i \rightarrow {\bf p}_i+m_i{\bf
v}_{f}^i$.\\
{\em iv) Coulomb propagation}\\
After break-up, hot primary fragments are supposed to suffer an
expansive propagation under their mutual Coulomb interaction. This stage can
be easily simulated by integrating the corresponding Newtonian
equations of motion. Integration has been carried out up to 500 fm/c, 
when Coulomb interaction between fragments can be neglected.\\
{\em v) Secondary decays}\\
Since primary fragments carry internal excitation, a secondary decay
stage was introduced in Ref. \cite{temp3}. Depending on the fragment
excitation, secondary break-up processes or particle evaporation were
considered. Since the above classification is rather arbitrary, here we
resume to the second process, treated using the standard Weisskopf
evaporation scheme. As in \cite{temp3}, the range of the evaporated
particles is considered up to $A=16$. Evaporation events are simulated
using standard Monte Carlo \cite{temp3}. 

For each primary break-up event (i), the steps ii) - v) are
performed. The resulted fragmentation events can be readily compared
with experimental ones (after removing the particles coming from the
preequilibrium stage).

\section{Comparison of the model predictions with recent
experimental data}

In this section comparison between the model results and the recent
experimental data concerning the reactions Xe+Sn at 32 MeV/u and Gd+U
at 36 MeV/u are presented. For the above reactions, various
experimental fragment size and kinematic observables have been
published recently \cite{indra1,indra2}.
%************** fig1.eps ***************************************** 
\begin{figure}
\includegraphics[height=9.5cm]{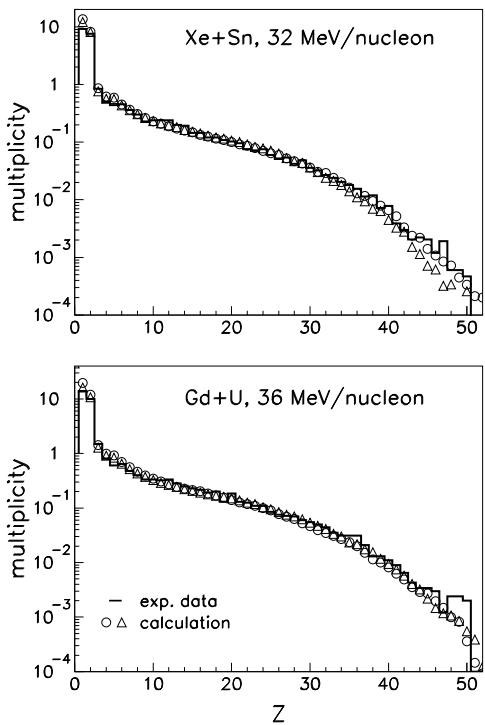}
\caption{Calculated charge multiplicities in comparison with
  experimental data for the Xe+Sn at 32 MeV/u and Gd+U at 36 MeV/u
  reactions. Open circles correspond to calculation using the
  freeze-out hypothesis (1), open triangles correspond to calculation
  using the hypothesis (2). Experimental data are represented by the
  histogram.}  \vspace*{-1cm}
\end{figure}
%*****************************************************************
The aim
of the study is to extract information about the physical phenomenon
taking place by trying to fit the entire set of experimental data
using the microcanonical multifragmentation model. 
It is known \cite{vfree} that various freeze-out assumptions may induce
important differences in the statistical models' results. For
drastically reducing this uncertainty in the present study we employ 
two opposite freeze-out scenarios:\\
1) The standard working hypothesis of the model: Fragments are
idealized as hard spheres placed into a spherical freeze-out
recipient; fragments are not allowed to overlap each other or the
recipient wall. The generation of the fragments' positions is described
in Ref. \cite{noi1}.\\
2) The hardcore interaction is switched
off: Integration over the fragments' positions may be approximately
carried out by assuming as in Ref. \cite{fai-ran} that each fragment
(from a configuration composed of $N$ fragments) is blocking the
volume $V_0/N$ ($V_0$ is the volume of the nuclear system at normal
density) for the rest of the fragments (as in \cite{fai-ran}) and for
itself as well. Coulomb energy is approximated by the Wigner Seitz
formula, being therefore independent of the fragments' positions
\footnote{For a $N$ fragment partition the Wigner Seitz Coulomb
  interaction energy writes:
$V_{WS}=\frac35\left(\frac{Z^2}R-\sum_{i=1}^N \frac{Z_i^2}{R_C^i}\right)e^2$
where $R$ is the radius of the freeze-out recipient and $R_C^i$ is the
radius of the Wigner Seitz cell corresponding to the fragment $i$,
$R_C^i=r_0(nA_i)^{1/3}$ where $r_0=1.2$ fm, and $n=V/V_0$.}.
The integration over the fragments' positions writes:
\begin{equation}
\int \prod_{i=1}^N {\rm d}{\bf r}_i  
\frac1{\sqrt{\det {\bf I}}} \prod_{j<i} \theta_{ij}
\simeq V_{free} \int \prod_{i=1}^N {\rm d}{\bf r}_i  
\frac1{\sqrt{\det {\bf I}}}
\label{eq:vf1}
\end{equation}
%************** fig2  ***************************************** 
\begin{figure}
\includegraphics[height=9.5cm]{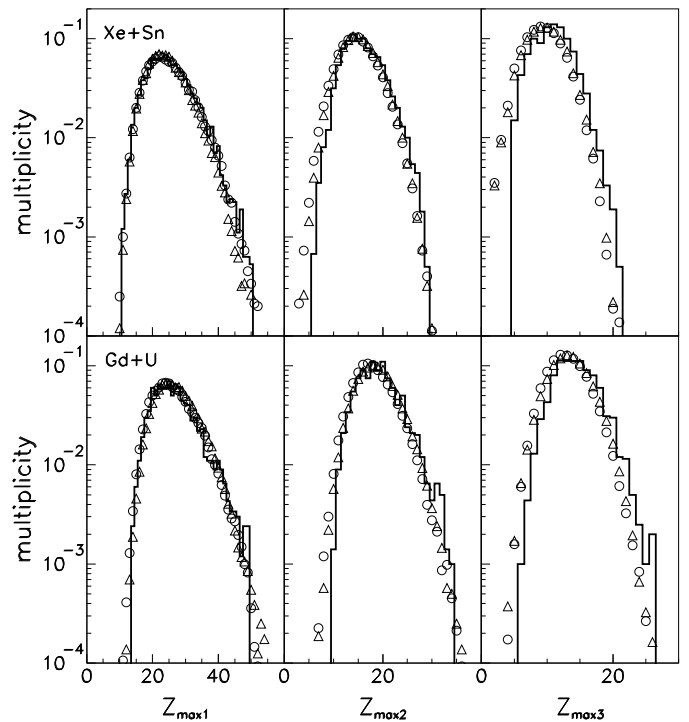}
\caption{Calculated distributions of the largest, second largest and third
  largest charge in one fragmentation event in comparison with the
  corresponding experimental data for the Xe+Sn at 32 MeV/u and Gd+U 
  at 36 MeV/u reactions. Symbols are used as in Fig. 1.\\[-.3cm]}
\end{figure}
%*****************************************************************
with:
\begin{equation}
V_{free}=\prod_{i=1}^N\left(V-i\frac{V_0}N\right).
\label{eq:vf2}
\end{equation}
The factor $\prod_{j<i} \theta_{ij}$ from eq. (\ref{eq:vf1}) is just a
formal expression of the fragment blocking constraint formulated
before.  With this approximation fragment positions have to be
generated into the spherical freeze-out volume, without any hardcore
constraint only for evaluating $\det {\bf I}$ and the statistical
weight of a configuration will get (as in Ref. \cite{vfree}) the extra
factor $V_{free}$. Having the weights of each configuration $C'$ the
corresponding Metropolis type simulation is straight forward \cite{noi1}.

Using the above defined working hypotheses, comparisons between the
model's results and the corresponding experimental data have been
performed. Both fragment size distributions and kinematic
distributions have been considered. Very good agreement between
calculations and experimental data can be observed for both reactions,
in both considered working hypotheses for all the considered
observables.

The comparisons between theoretical and experimental fragment size
distributions corresponding to both Xe+Sn at 32 MeV/u and Gd+U at 36
MeV/u reactions are as follows: Charge multiplicities are given in
Fig. 1. Multiplicities of largest charge ($Z_{max1}$), second largest
charge ($Z_{max2}$) and third largest charge ($Z_{max3}$) from one
fragmentation event are presented in Fig. 2.
%************** fig3  ***************************************** 
\begin{figure}
\includegraphics[height=8cm]{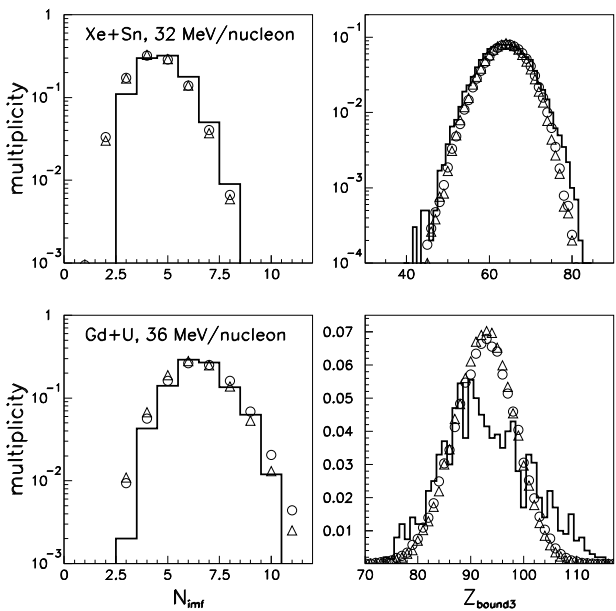}
\caption{Calculated IMF multiplicities (left column) and $Z_{bound3}$
  multiplicities (right column) in comparison with the corresponding
  experimental data for the reactions Xe+Sn at 32 MeV/u (upper raw)
  and Gd+U at 36 MeV/u (lower raw). Symbols are as in Fig.
  1.\\[-.8cm]}
\end{figure}
%*****************************************************************
In Fig. 3 intermediate mass fragment (IMF) distributions and
$Z_{bound3}$ distributions are given (IMF fragments are considered to
have $Z \ge 5$ as in \cite{indra2}; $Z_{bound3}$ is defined as the sum
over all fragments' atomic numbers from one fragmentation event which
are greater or equal to 3; $Z_{bound5}$ is similarly defined but the
$Z$ limit is fixed to 5). $Z_{bound5}$ multiplicities are compared
with experimental data in Fig. 4 only for the reaction Gd+U at 36
MeV/u. It can be noticed that for the $Z_{bound}$ distributions the
calculated width of the distributions are slightly smaller than
the experimental ones. This suggests a (small) fluctuation of the
experimental equilibrated source. 
%************** fig4  ***************************************** 
\begin{figure}[b]
\includegraphics[height=5.5cm]{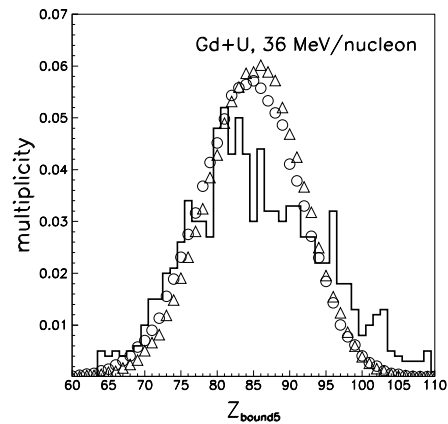}
\caption{Same as in Fig. 3 (right column) for the $Z_{bound5}$
  distribution corresponding to the Gd+U at 36 MeV/u reaction.\\[-.6cm]}
\end{figure}
%*****************************************************************
This will become more clear in the next section. An excellent agreement
for all considered observables related to fragment size was,
therefore, obtained.

Comparisons between theoretical and experimental average fragment
kinetic energy function of fragment charge are given in Fig. 5, left
column for the the two considered reactions. Very good agreements are
to be noticed. A deeper insight to the kinematics of the process can
be obtained by analyzing the reduced fragments velocity correlations
\cite{gross}. Fragments with $5 \le Z \le 20$ as in
\cite{indra-bologna} have been used for constructing the correlations.
This observable was proven to reflect subtle topological features of
the fragmentation events. The beginning of the correlation function is
often called ``Coulomb hole'' and reflects the strength of the Coulomb
repulsion between fragments (obviously related to the size of the
freeze-out volume) but, as shown in the next section, is also related
to the amount of radial expansion.  Good agreement between
calculations and experimental data can be observed here as well (see
Fig. 5). While in the case of the Gd+U reaction the experimental
velocity correlations are well reproduced in both freeze-out
hypotheses, in the case of the first reaction a better agreement with
experimental data is obtained for the hypothesis (2) (``without
hard-core'').
%************** fig5  ***************************************** 
\begin{figure}
\includegraphics[height=8cm]{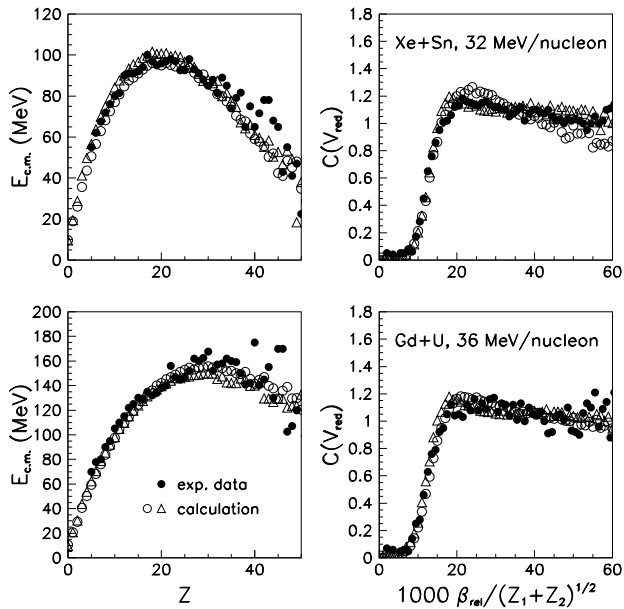}
\caption{Average fragment kinetic energy versus charge (left column)
  and reduced fragment velocity correlation (right column) for the
  reactions Xe+Sn at 32 MeV/u (upper raw) and Gd+U at 36 MeV/u (lower
  raw). Symbols are as in Fig. 1 except for the experimental data
  which are here represented by full circles.\\[-.6cm]}
\end{figure}
%*****************************************************************
In fact, this result is natural and suggests that deviations from the
spherical fragments with hard-core interaction idealization (deformed
fragments, surface diffusivity, etc.) are present in real
multifragmentation. The deviation effect is stronger in the case of
the Xe+Sn since the corresponding equilibrated system is smaller and,
therefore the fragment partitions are more symmetric (as shown in
\cite{vfree}, the free volume is smaller for more symmetric partitions
and larger for more asymmetric ones). The simultaneous description of
both average fragment kinetic energies as a function of the fragment
charge and of the reduced velocity correlations is remarkable and
reflects a good description of the event topology.  Here it is worth
noticing that in Ref. \cite{indra-bologna} the velocity correlations 
for both reactions could not be reproduced via the SMM model.
%************** fig6  ***************************************** 
\begin{figure}[b]
\includegraphics[height=6.5cm]{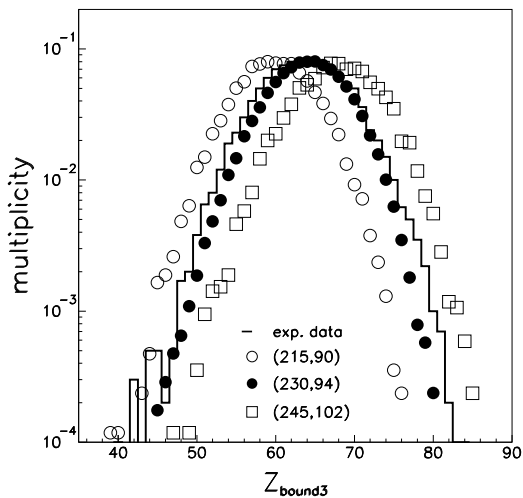}
\caption{$Z_{bound3}$ distributions (symbols) corresponding to
  various sizes of the nuclear source within the working hypothesis
  (1) (see the legend).\\[-.7cm]}
\end{figure}
%*****************************************************************

The values of the model input parameters for which the comparisons
with the experimental data have been performed are listed in Table
\ref{tab:1}. A remarkable fact is that the fitting parameters
corresponding to the freeze-out hypotheses (1) and (2) are quite close
ones to the others for each of the considered reactions.
%, in spite of the large difference between the hypotheses. 
This tends to indicate a
relatively small ``region'', specific to each reaction in which the
parameters of the {\em real} equilibrated nuclear source are
situated. Quite small differences from one hypothesis to the other
seem to be present in the case of the freeze-out volume. This result
%is quite surprising 
may look intriguing
given the fact that freeze-out volume is directly
related to both working hypotheses. In reality, the
relatively large freeze-out volumes obtained are minimizing the
influence of the hard-core interactions to the free-volume
($V_{free}$) (see e.g. \cite{vfree}), 
%%%%%%%%%%%%%%%%%%%%%%%%%%%%%%%%%%%%%
the values of of the freeze-out volume being mainly dictated  
by the Coulomb interaction (the remaining quantity entering the
system's density of states depending on the freeze-out volume). 
Unlike the free volume which in models can only be more or less
arbitrary parametrized, Coulomb interaction can be accurately 
evaluated for each fragment partition (either by explicitly accounting 
for the interfragment interaction, or by employing the Wigner-Seitz
approach). This contributes to the solidity of the present evaluations. 
%[It is worth mentioning that similar values of $V$ have
%been obtained in the framework of the Brownian One Body
%dynamics (BoB) model (e.g. $V=8 V_0$ for the Gd+U reaction). \cite{Borderie}] 
 
%%%%%%%%%%%%%%%%%%%%%%%%%%%%%%%%%%%%% 
The differences in the
excitation energies corresponding to the two hypotheses are due to the
Coulomb energy variation from one case to another, caused by the slightly
different values of the freeze-out volumes (see Table
\ref{tab:1}). For example, in the Gd+U case the energy variation (from
(1) to (2)) is slightly larger than in the Xe+Sn case since the
{\em relative} variation of the freeze out volume is larger as
well. 
Freeze-out volume is always smaller in the working hypothesis
(2). This is related to the different dependencies $V_{free}(V)$
resulting from the underlying hypotheses (more symmetric partitions
and thus a more advanced fragmentation are specific to the without
hardcore case and therefore both the freeze-out volume and excitation
energies need to be smaller in the second hypothesis). The flow
parameters were chosen such as to insure a good reproduction of the
kinetic observables (see Fig. 5). The way in which each parameter is
influencing the fitting observables is discussed in the next
section. 
\begin{table}
\caption{\label{tab:1} Values of the model's input parameters
corresponding to the fitting of the experimental data for both Xe+Sn
at 32 MeV/u and Gd+U at 36 MeV/u reactions in the freeze-out
hypotheses (1) and (2). The last two columns correspond respectively
to the flow energy and to the exponent defining the flow profile (see
text).\\[.1cm]}
\begin{ruledtabular}
\begin{tabular}{rccccc}
react./hyp.&$(A,Z)$&$V/V_0$&$E_{ex}$(MeV/u)&$E_{fl}$(MeV/u)& $\alpha$\\
\hline
Xe+Sn (1)&(230,94)&9&5.3&1.4&2\\
Xe+Sn (2)&(220,92)&8.5&4.78&1.9&1.8\\
\hline
Gd+U (1)&(343,136)&8.5&5.95&1.7&2\\
Gd+U (2)&(328,130)&8&5.2&1.9&1.8
\end{tabular}
\end{ruledtabular}
\end{table}

\section{Influence of the model parameters on various fitting observables}

While many fragment size and kinetic observables were simultaneously
fitted in order to deduce the parameters of the equilibrated sources
corresponding to the two reactions, the following question arises: are
they unique? We address this question in the present section.

We start with the remark that the size of the equilibrated source
appears to be dictated by the $Z_{bound}$ distribution. We illustrate
this by representing in Fig. 6 the $Z_{bound3}$ distributions
corresponding to three different sources: (215,90), (230,94) and
(245,102). 
%************** fig7  ***************************************** 
\begin{figure*}
\includegraphics[height=8cm]{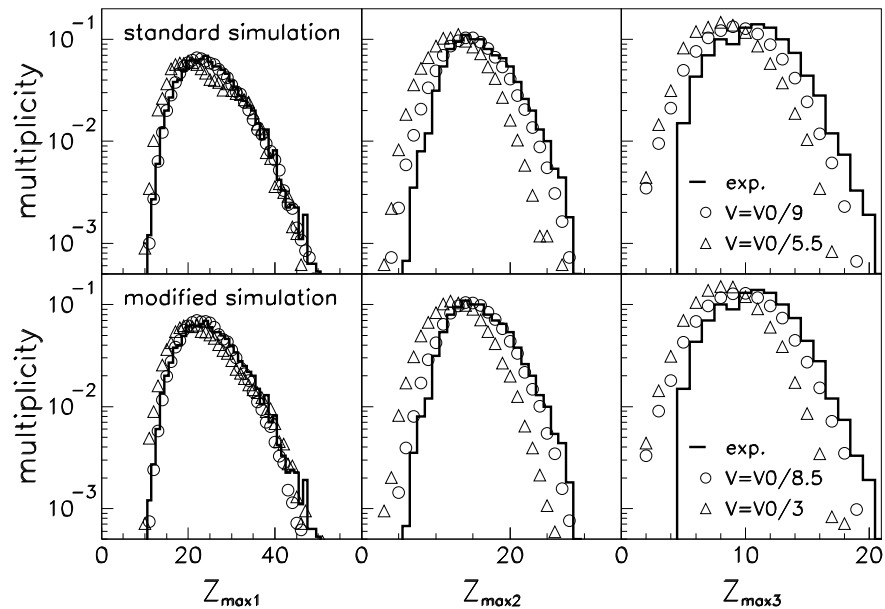}
\caption{Largest, second largest and third largest charge
  distributions corresponding to two different freeze-out volumes (see
  the legend) for each of the two freeze-out hypotheses ((1) upper raw;
  (2) lower raw) in comparison with Xe+Sn at 32 MeV/u the experimental
  data. See the text for the sources' parameters.}
\end{figure*}
%*****************************************************************
%%%%%%%%%%%%%%%%%%%%
Their excitation energies (5.4, 5.3, 5.3 MeV/u respectively) are
chosen such as to insure a good fit of the charge distributions
corresponding to the Xe+Sn at 32 MeV/u reaction; their freeze-out
volume ($V/V_0=9$) is chosen as to provide a good description of
the $Z_{max}$, $Z_{max2}$ and $Z_{max3}$ distributions.
%%%%%%%%%%%%%%%%%%%%%%%%%%%%%%%%%%%%%%%%%%%%%%%%%%%%%
It can be easily observed that the best fit on the data is
given by the (230,94) source. A smaller source deviates the maximum of
the $Z_{bound3}$ distribution to the left while a larger one deviates
it to the right (see Fig. 6).

The freeze-out volume of the equilibrated sources appears to be
dictated by the distributions of the largest, second
largest and third largest charge in
one fragmentation event. Obviously, these distributions provide a
measure of the degree of asymmetry of a given fragment partition. The
degree of asymmetry of a partition is influenced by the freeze-out
volume through both the Coulomb interaction and the free volume
\cite{vfree}. This fact is evidenced in Fig. 7 where multiplicities of
$Z_{max1}$, $Z_{max2}$ and $Z_{max3}$ are represented for two different
freeze out volumes in the freeze-out hypotheses (1) and (2) as
follows. For the first hypothesis the source was 
considered (230,94) at two different volumes: $V_0/5.5$ and $V_0/9$
while for the second one the source was taken to be (220,92) at other
two different volumes $V_0/3$ and $V_0/8.5$. The corresponding
excitation energies are taken such as a good reproduction of the
experimental charge distributions corresponding to the Xe+Sn reaction
at 32 MeV/u to be achieved (the excitations energies are respectively:
6.4, 5.3, 6.1, 4.78 MeV/u). It can be noticed that deviations to the
left of the maxima (indicating more asymmetric fragment partitions) 
of the  $Z_{max2}$ and $Z_{max3}$ distributions are occurring when the 
freeze-out volume is decreased. 
%************** fig8  ***************************************** 
\begin{figure}[b]
\includegraphics[height=8cm]{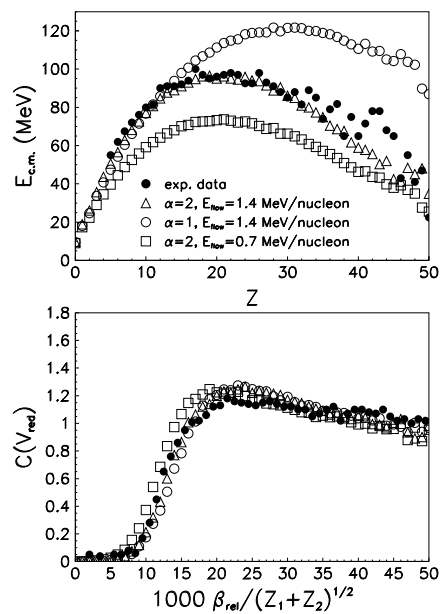}
\caption{Influence of the radial flow parameters on the fragment average
  kinetic energy versus charge dependency and on the fragment reduced
  velocity correlations for the working hypothesis (1) (see the legend
  and the text for the sources' parameters). The experimental data are
  here represented by full circles.}
\end{figure}
%*****************************************************************
%%%%%%%%%%%%%%%%%%%%%%%%%%%%%%%%
This behavior can be observed irrespective to the considered
freeze-out hypothesis or to the chosen mass of the source. 
%%%%%%%%%%%%%%%%%%%%%%%%%%%%%%%%
While, as demonstrated by Fig. 7 lower values of the freeze-out volume are
leading to deviations of the $Z_{max1}$, $Z_{max2}$ and $Z_{max3}$
calculated distributions from the corresponding experimental data,
larger values of the freeze-out volume will result in smaller
excitation energies and larger flow energies leading to deviations
from the experimental data of the kinematic observables represented in
Fig. 5. 

%%%%%%%%%%%%%%%%%%%%%%%%%%%%%
Given the previously described monotonical behavior of the
``distance'' from the calculated curves to the experimental ones, and
also the very good reproduction of all the considered experimental
observables for the parameters (given in Table I) corresponding to the
case in which the $Z_{bound3}$ distribution is fitted best (see Fig.
6) it results that the corresponding sets of parameters provide the
global minimum of the associated $\chi^2$ function. Indeed, choosing a
different value of $A$ would generate a deviation of the $Z_{bound3}$
from the experimental one and, obviously the global reproduction of
the data would be worst.
%%%%%%%%%%%%%%%%%%%%%%%%

Finally, the way in which the radial flow parameters $E_{fl}$ and
$\alpha$ are influencing the experimental data fitting is presented in
Fig. 8 for the nuclear source (230,94), $V=V_0/9$, $E_{ex}=5.3$ MeV/u and
the first working freeze-out hypothesis. Three different cases are
being considered: ($\alpha=2$, $E_{fl}=1.4$ MeV/u), ($\alpha=1$,
$E_{fl}=1.4$ MeV/u) and ($\alpha=2$, $E_{fl}=0.7$ MeV/u). As shown in
Fig. 8 the $\alpha$ parameter defining the flow profile appears to
influence the location of the maximum of the average kinetic energy
versus $Z$ ($E_{c.m.}(Z)$) and the value of the flow energy the height
of the $E_{c.m.}(Z)$ dependency. Smaller effects are to be noticed in
the reduced velocity correlation function and they concern mainly the
width of the ``Coulomb hole''. The effects are intuitive: smaller flow
leads to a narrower Coulomb hole while larger ones lead to a larger
hole. The parameters $\alpha=2$ and $E_{fl}=1.4$ MeV/u appear to give
the best description of the kinetic experimental data corresponding
to the Xe+Sn at 36 MeV/u reaction. 

Therefore, it results that, within some (small) inherent
uncertainties, the obtained parameters corresponding to the two
working hypotheses ((1) and (2)) are uniquely determined.

\section{Conclusions}

In summary, the microcanonical multifragmentation model \cite{noi1}
has been further refined and improved. Conservation of angular
momentum has been included. Also, in order to provide events similar
with the experimental ones, the model is completed with a stage of
generation of momenta for the primary break-up fragments. Radial flow
velocities can be superimposed here. Finally, an expansion stage of
the hot primary fragments under their mutual Coulomb interaction is
added. Using the new version of the model, the recent experimental
data of the INDRA collaboration concerning the reactions Xe+Sn at 32
MeV/u and Gd+U at 36 MeV/u are analyzed. Both fragment size
distribution observables and kinematic observables could be
simultaneously fitted allowing thus the identification of the
equilibrated sources parameters corresponding to the two reactions.
Two 
%opposite 
different break-up scenarios were used for performing the analysis.
The two types of calculations give close results for each of the
considered reactions suggesting independence of the obtained results
on the particular hypothesis.
%%%%%%%%%%%%%%%%%%%%%%%%%%%%%%%%%%%%%%%%
This is due to the fact that
for relatively large values of the freeze-out volume (such as those 
obtained by fitting the data, 8-9 $V_0$) the influence of the
hard-core is small and thus the results are rather independent on
the extent to which these effects are included in a specific
$V_{free}$ parametrization. In this case, the values of the
freeze-out volume are dictated by the Coulomb interaction (accurately 
evaluated independently to any model assumptions) which makes 
the evaluation particularly robust. For each reaction a narrow 
freeze-out volume region was identified (8.5-9
$V_0$ for Xe+Sn and 8-8.5 $V_0$ for Gd+U). 
%%%%%%%%%%%%%%%%%%%%%%%%%%%%%%%%%%%%%%%%%%%%%%
%This unambiguously indicates a
%relatively small region corresponding to each reaction where the {\em
%real} parameters of the equilibrated sources are situated. 
By analyzing the way in which variations of the model parameters are
influencing the deviation between the calculated curves and the
experimental ones, it is proved that the set of the parameters
providing the best fit to the data is unique (within inherent
uncertainties).  It is shown that the size of
the source is dictated by the $Z_{bound}$ distributions while its
freeze-out radius can be deduced by fitting the experimental
distributions of $Z_{max1}$, $Z_{max2}$ and $Z_{max3}$ and the
kinematic observables. The way in which the flow energy and the
parameter $\alpha$, defining the flow profile is influencing the
fragment average kinetic energy versus $Z$ and also the reduced
velocity correlation is also discussed.  The above parameters are
shown to be clearly identifiable as well.  To our knowledge, this is
the first complete reproduction of an entire set of experimental data
(including both fragment size distribution observables and kinematic
observables) by means of a microcanonical multifragmentation model.
Moreover, it was shown for the first time that the parameters of the
experimentally obtained equilibrated source can be uniquely determined
by means of such a model.
%Apart from the fact that it proves the possibility
%of identifying the freeze-out parameters of the statistically
%equilibrated systems formed in heavy ions collision, the present
%analysis identifies for the first time a parameter crucial for
%establishing the thermodynamics taking place in such systems: the
%freeze-out volume. It is remarkable that a narrow freeze-out volume
%region is indicated irrespective to the freeze-out hypothesis. This
%proves that the obtained values are rather independent to the various
%model hypotheses.
\\[-.8cm]
\begin{acknowledgments}
The authors acknowledge the support of the Alexander von Humboldt
Foundation during this work. 
\end{acknowledgments}

\end{document}